\documentclass[aps,prl,reprint,amsmath,amssymb,
superscriptaddress,showpacs,showkeys]{revtex4-1}
\usepackage{graphicx}
\usepackage{dcolumn}
\usepackage{bm}
\usepackage{soul} 
\usepackage{color}
\definecolor{vs}{rgb}{0.1,0.4,0.1}                  

\newcommand{\del}[1]{}                             
\usepackage{hyperref}
\usepackage{verbatim}
\newcommand\wordcount{\verbatiminput{\jobname.sum}}
\begin{document}


\title{Paradox of multiple plasmonic resonances at light scattering by a cylinder of infinitesimal radius}


\author{Yaroslav A. Brynkin}
\affiliation{
M. V. Lomonosov Moscow State University, Moscow, 119991, Russia}
\author{Michael I. Tribelsky}
\email[Corresponding author (replace ``\_at\_" by @):\\]{\mbox{E-mail: mitribelsky\_at\_gmail.com}
}
\homepage[]{https://polly.phys.msu.ru/en/labs/Tribelsky/}
\affiliation{
M. V. Lomonosov Moscow State University, Moscow, 119991, Russia}
\affiliation{National Research Nuclear University MEPhI (Moscow Engineering Physics Institute), Moscow, 115409, Russia}
\affiliation{RITS Yamaguchi University, Yamaguchi, 753-8511, Japan}

\date{\today}

\begin{abstract}
The paradox of the divergence of the resonant scattering cross section of a cylinder with the permittivity equals minus unity and vanishing radius (R) irradiated by a monochromatic electromagnetic wave is discussed. Within the framework of the exact solution of the Maxwell equations, the divergence at the specified conditions is caused by the overlap of all but one multipolar resonances. It is shown that the paradox is caused by the too straightforward analysis of the expression for the cross section, which has a singularity at this point. To resolve the singularity, one must, first, generalize the problem formulation, taking into account the final linewidth of the incident wave, and then perform the correct sequence of limit transitions. The application of this approach gives rise to the vanishing cross section at the vanishing R. It ruins the expectations to employ such a cylinder as a superscatterer but simultaneously open a door to counterintuitive effects both in far and near field zones related to unusual size dependences of the scattered fields at small but finite R.
\end{abstract}

\maketitle
%
%
\emph{Introduction.} Despite more than hundred years history of the study of light scattering by obstacles the problem still remains among the most important issues of electrodynamics. In addition to the purely academic interest (plenty of new effects have been discovered recently, see, e.g., the discussions in Refs.~\cite{Grigorenko2005,Miroshnichenko:RMP:2010}) it is explained by the great demand from numerous technologies, ranging from medicine and biology~\cite{schwartz2009feasibility,Tribelsky2016} to systems for data storage and processing and telecommunications~\cite{Tribelsky2015}.

Nowadays, the frontier of this study shifts to the nanoscale~\cite{Klimov:Book:2014}. In a number of applications, e.g., when nanoparticles are used as biomarkers, it important to enhance their scattering cross sections, since usually, they are small owing to the small geometric sizes of the particles. It stimulates quests of a \emph{superscatterer}~\cite{Ruan2010}. The main idea of it is to employ a resonant scattering instead of non-resonant and to manage overlap of several resonances.

Then, it seems, the best way to design the superscatterer is to employ the plasmonic resonances at light scattering by a low-lossy cylinder of a small radius. The physical grounds for this are the following: When the dissipative losses in an illuminated subwavelength particle are so small that the radiative ones begin to prevail, the Rayleigh scattering is replaced by anomalous~\cite{Tribelsky2006}.
At the anomalous scattering, the corresponding partial cross section achieves its theoretical upper bound, which does not depend on the particle size. For a spatially-uniform spherical subwavelength particle of the infinitesimal radius illuminated in a vacuum the resonant value of the permittivity $\varepsilon$, for the $\ell$th partial mode is \mbox{$-(\ell+1)/\ell$}, where \mbox{$\ell = 1,\;2\ldots$} designates the dipole, quadrupole, etc. Through the dispersion relation $\varepsilon(\omega)$, where $\omega$ stands for the frequency of the incident wave, it determines the resonant frequencies $\omega_\ell$. Importantly, that, generally speaking, all $\omega_\ell$ are different and overlap of the resonances of different orders does not take place. To achieve the overlap more sophisticated particle structure, such as, e.g., core-shell is required~\cite{Ruan2010}.

This is not the case for a cylinder. The problem for the scattering of a plane, linearly polarized electromagnetic wave by a right circular infinite spatially-uniform cylinder of radius $R$ is exactly solvable for any $R$ and any orientations of the cylinder axis with respect to the wave vector of the incident wave $\mathbf{k}$ and its polarization plane, see, e.g. book~\cite{bohren2008absorption}. The solution has a form of the infinite series of the multipolar contributions \mbox{$(-\infty < \ell < \infty)$.} However, the dramatic difference with the spherical particle is that now at $R \rightarrow 0$ the resonant $\varepsilon$  for all multipoles but the one with $\ell = 0$ converge to \emph{one and the same value} $\varepsilon = -1$, see, e.g., Ref.~\cite{lukyanchuk2007peculiarities931101}. Since at the anomalous scattering the resonant partial cross section (per unit length of the cylinder) $\sigma_{\rm sca}^{(\ell)}$ is a finite quantity which does not depend on $R$ and, therefore, remains finite at $R \rightarrow 0$, see below, the overlap of the \emph{infinite number} of resonances means \emph{divergence} of the multipolar series.

Thus, we face a paradox: On the one hand, it is quite obvious, that at $R \rightarrow 0$ the scattering should be suppressed. On the other hand, the \emph{exact} solution gives rise to the infinitely-large scattering cross section. The argument that the divergence occurs only in the non-dissipative limit and that any finite dissipation cuts the divergence off does not help much --- the fundamental principles of electrodynamics do not provide any limit for the lower bound of the dissipation at a given $\omega$. Then, at low enough Im$\,\varepsilon$ a cylinder with the radius of several nanometers may have the scattering cross section of a few square kilometers, which also is nonsense.
The goal of our paper is to resolve the paradox and to answer the question, if the multiple resonances in the plasmonic cylinder may be employed to create a simple subwavelength superscatterer.

\emph{Problem Formulation.} To understand the physical grounds for the paradox note that it is associated with the following reasons: First, the dependencies $\sigma_\ell(\varepsilon,\,R)$ do not have any definite limit at the point $\varepsilon = -1,\;R = 0$ --- their values depend on the shape of the trajectory in the plane $(\varepsilon,\,R)$ along which one approaches this point, see below. Second, any actual light source always has a finite linewidth and, therefore, a certain spectral distribution of its power. On the other side, at the anomalous scattering, the resonant linewidths rapidly vanish at $R \rightarrow 0$~\cite{Tribelsky2006,lukyanchuk2007peculiarities931101}. Then, at any finite but fixed source linewidth at small enough $R$ only a certain part of the light power is scattered resonantly. For the accurate problem treatment, these two issues should be taken into account explicitly.

For the sake of simplicity, we consider the normal incidence of the TE-polarized plane wave so that vector $\mathbf{k}$ is perpendicular to the axis of the cylinder and vector $\mathbf{E}$ oscillates in the plane of the base of the cylinder. The extension of our consideration to the general case of an arbitrary incidence is quite straightforward. It gives rise to the same conclusions but requires more cumbersome calculations.

\emph{Analysis.} The exact solution of the formulated problem is presented as the infinite series in the cylindrical functions. This structure of the solution is universal and valid for any $\varepsilon$. However, to consider the paradox in its extreme form, in what follows we discuss the non-dissipative case, i.e., $\varepsilon$ is supposed to be purely real.

All specific information about the scattering for a given pair $(\varepsilon,\,R)$ is hidden in the coefficients of this series, called the \emph{scattering coefficients}. For the problem in question the scattering field outside the cylinder is described by the single set of the scattering coefficients
\begin{equation}\label{eq:a_ell}
{a_\ell} = \frac{{{\mathfrak{F}_\ell}}}{{{\mathfrak{F}_\ell} + i{\mathfrak{G}_\ell}}},
\end{equation}
where
\begin{equation}\label{eq:F_ell_G_ell}
  \begin{gathered}
  {\mathfrak{F}_\ell} = m{J_\ell}\left( {mq} \right){J'_\ell}\left( q \right) - {J'_\ell}\left( {mq} \right){J_l}\left( q \right), \hfill \\
  {\mathfrak{G}_\ell} = m{J_\ell}\left( {mq} \right){N'_\ell}\left( q \right) - {J'_\ell}\left( {mq} \right){N_\ell}\left( q \right), \hfill \\
\end{gathered}
\end{equation}
and $J_\ell(z),\; N_\ell(z)$ stand for the Bessel and Neumann functions, respectively, the prime indicates differentiation with respect to the entire argument of the corresponding function, $m \equiv \sqrt{\varepsilon}$ is the refractive index~\cite{bohren2008absorption}, the size parameter $q$ equals $kR$, \mbox{$k \equiv |\mathbf{k}| \equiv \omega/c$,} denotes the wavenumber of the incident wave, $c$ stands for the speed of light in a vacuum.

In terms of $a_\ell$ the partial ($ \sigma_{\rm sca}^{(\ell)}$) and the net  $(\sigma_{\rm sca}^{\rm net}$) scattering cross sections are expressed in a very simple manner~\cite{bohren2008absorption}:
\begin{equation}\label{sigma_omega}
 \sigma_{\rm sca}^{(\ell)} = \frac{4\pi}{k} {{{\left| {{a_\ell}} \right|}^2}},\; \sigma_{\rm sca}^{\rm net} = \sum\limits_{\ell =  - \infty }^\infty \sigma_{\rm sca}^{(\ell)} \equiv \sigma_{\rm sca}^{(0)} + 2 \sum\limits_{\ell =  1 }^\infty \sigma_{\rm sca}^{(\ell)},
\end{equation}
the latter owing to the identity $a_\ell \equiv a_{-\ell}$.

We are interested in the limit $q \rightarrow 0$. Then, the expansion of the functions in the right-hand-side of Eq.~\eqref{eq:F_ell_G_ell} in powers of small $q$ yields~\cite{lukyanchuk2007peculiarities931101}

\begin{eqnarray}
&\hspace*{-2cm}{\mathfrak{F}_\ell}& \cong
\left\{
  \begin{array}{ll}
{\displaystyle\frac{m(\varepsilon -1)}{16}}q^3+\ldots\;\; & \textrm{at \mbox{$\ell=0$}} \vspace*{5pt}\\
{\displaystyle\frac{(\varepsilon -1) m^{\ell -1} }{2^{2\ell}\ell!(\ell-1)!}}q^{2 \ell -1}+\ldots & \textrm{at \mbox{$\ell \neq 0$}}\vspace*{5pt}\\
\end{array}\right.\label{eq:F_ell_q}\\
&\hspace*{-2cm}{\mathfrak{G}_\ell}& \cong
\left\{
  \begin{array}{ll}
{\displaystyle\frac{2m}{\pi q}}+\ldots & \textrm{\hspace*{-3.68cm}at \mbox{$\ell=0$}},\vspace*{5pt}\\
{\displaystyle\frac{\varepsilon +1}{\pi q}-\frac{q}{8\pi}(\varepsilon -1)\left[2+\varepsilon-4\log\frac{qC}{2}\right]}+\ldots \\&\textrm{\hspace*{-3.68cm}at \mbox{$\ell=1$}} \vspace*{5pt}\\
{\displaystyle\frac{m^{\ell -1}(\varepsilon +1)}{\pi q} -\frac{q}{4\pi}(\varepsilon -1)\left[\frac{\varepsilon}{\ell+1}+\frac{1}{\ell-1}\right] +\ldots}\vspace*{5pt}\\
 &\textrm{\hspace*{-3.68cm}at \mbox{$\ell>1$}}\\
\end{array}\right.\label{eq:G_ell_q}
\end{eqnarray}
Here $C\equiv \exp(\gamma)=1.78107\ldots$ and $\gamma = 0.577...$ stands for Euler’s constant.

The zeros of $\mathfrak{G_\ell}$ correspond to the Mie resonances. At the resonances $|a_\ell|$ achieves its the maximal value equals unity, see Eq.~\eqref{eq:a_ell}. Thus, equation  $\mathfrak{G}_\ell(\varepsilon,q) =0$ determines the resonance trajectories. Since along the trajectories \mbox{$a_\ell=1$} and all trajectories but the one with $\ell=0$ at $q \rightarrow 0$ converge to the same point with $\varepsilon=-1$, see Eq.~\eqref{eq:G_ell_q}, the aforementioned paradox becomes well-pronounced.

The question we are interested in is the linewidth of the $\ell$-th order resonance. As it is seen from Eq~\eqref{eq:G_ell_q}, the mode with $\ell=0$ does not have the resonance in the vicinity of the point \mbox{$q=0$}, \mbox{$\varepsilon =-1$}. Therefore, in what follows, discussing the scattering coefficients and the partial cross sections it is tacitly accepted that $\ell \neq 0$.

Note now that, actually, as it has been already mentioned above, the coefficients $a_\ell$ at the point \mbox{$q=0,\;\varepsilon =-1$} do not have any definite value at all. To show this, first, cancel the common factor $m^{\ell -1}$ in the numerator and denominator of Eq.~\eqref{eq:a_ell}. Then, $\mathfrak{F}_\ell$ and $\mathfrak{G}_\ell$ transform into purely real $F_\ell$ and $G_\ell$, respectively~\footnote{The possibility to transform $\mathfrak{F}_\ell$ and $\mathfrak{G}_\ell$ in Eq.~\eqref{eq:a_ell} into purely real $F_\ell$ and $G_\ell$ at purely real $\varepsilon$ follows from Eq.~\eqref{eq:F_ell_G_ell} and is not related to the smallness of $q$.}. Next, invert the relations \mbox{$F_\ell = F_\ell(q,q\sqrt{\varepsilon})$,} $G_\ell = G_\ell(q,q\sqrt{\varepsilon})$ considering $F_\ell$ and $G_\ell$ as new independent variables. Then, the plane $(\varepsilon,q)$ maps into the plane $(F_\ell,G_\ell)$ and the point \mbox{$\varepsilon =-1,\; q=0$} maps into the point (0,0).

To evaluate the indeterminate form of the type 0/0, which Eq.~\eqref{eq:a_ell} is transformed to at this point, let us consider a trajectory in the $(F_\ell,G_\ell)$ plane approaching the coordinate origin. Without loss of generality we can suppose that in the vicinity of the origin the trajectory is described by the expression $G=AF^\alpha$, where $A$ and $\alpha$ are any real constants. Then, it is trivial to see from Eq.~\eqref{eq:a_ell} that at $F \rightarrow 0$ coefficient \mbox{$a_\ell \rightarrow 1$} at \mbox{$\alpha>1$}; \mbox{$a_\ell \rightarrow 0$} at \mbox{$\alpha<1$}; and  \mbox{$a_\ell \rightarrow 1/(1+iA)$} at \mbox{$\alpha=1$}.

\begin{figure}
  \centering
  \includegraphics[width=.5\textwidth]{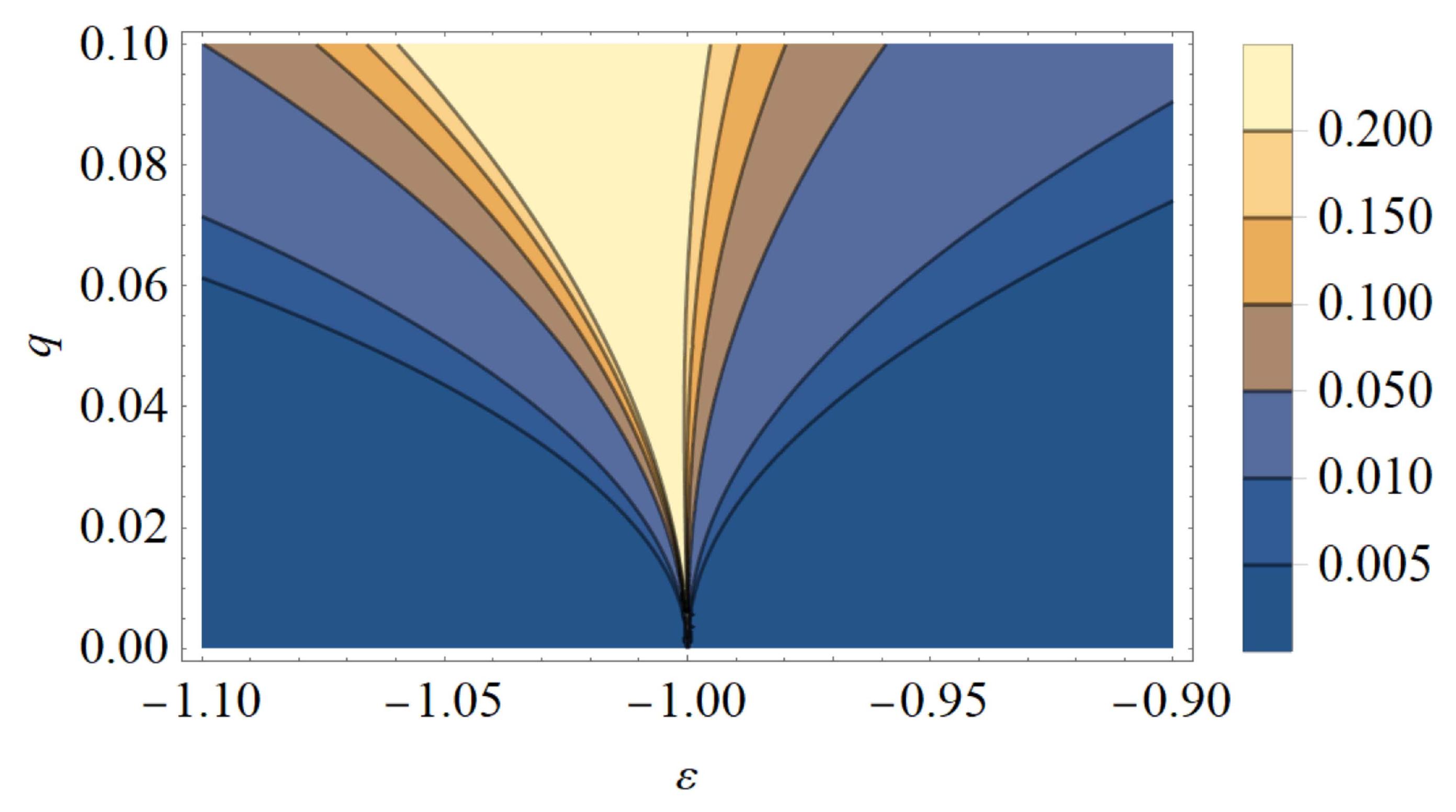}
  \caption{Contour plot of $|a_1(\varepsilon,q)|^2$ in the vicinity of the singular point (-1,0). Note, all contours corresponding to different values of $|a_1|^2$ merge at the same singular point. See the text for details. }\label{fig:F1}
\end{figure}

\begin{figure}
  \centering
  \includegraphics[width=.44\textwidth]{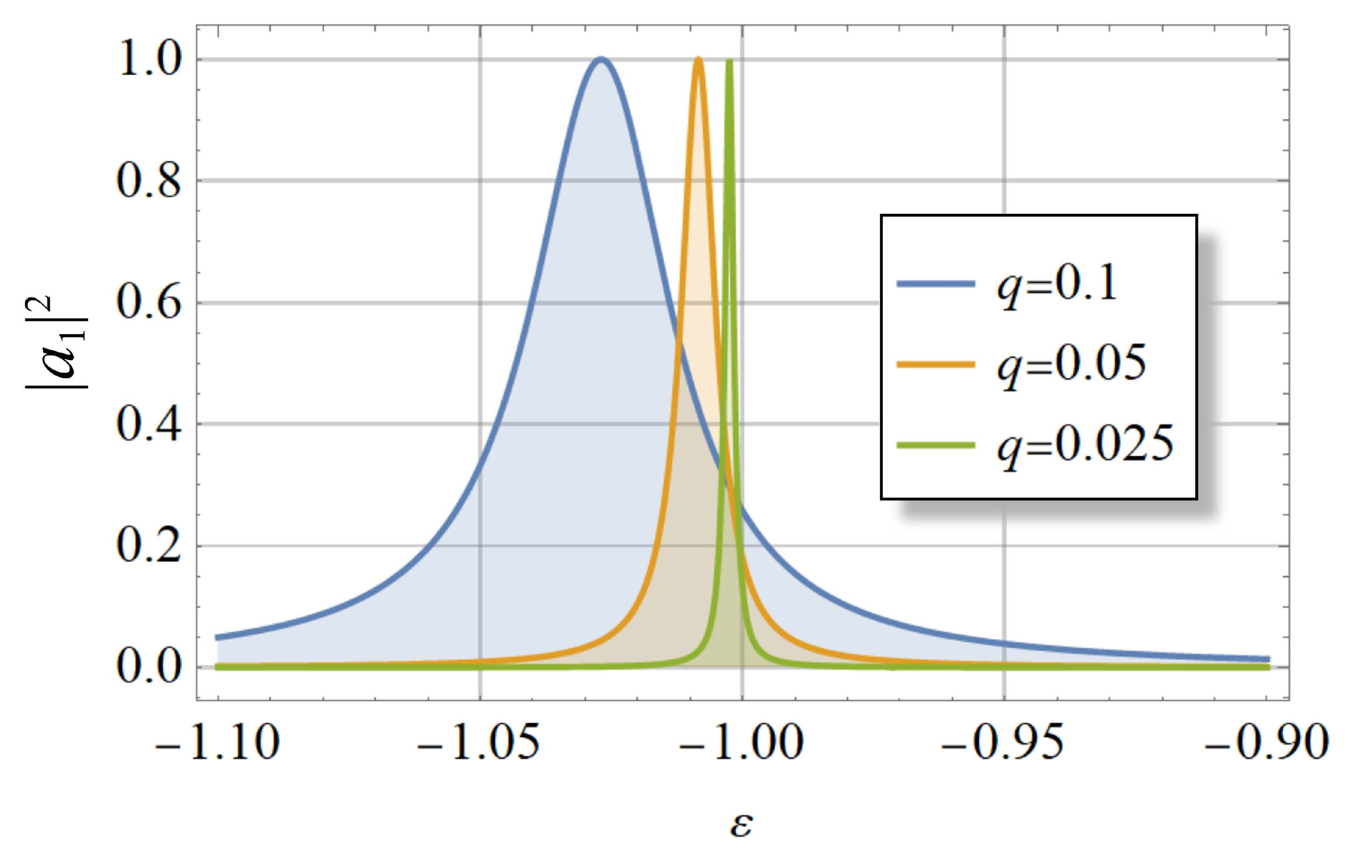}
  \caption{Sections of the profile $|a_1(\varepsilon,q)|^2$ by the planes \mbox{$q=0.1$}; \mbox{$q=0.05$} and \mbox{$q=0.025$}. See the text for details. }\label{fig:F2}
\end{figure}

To reveal the grounds for such an unusual behavior of the seemingly smooth function, let us consider the plots $|a_\ell(\varepsilon,q)|^2$ at the vicinity of the specified singular point. All of them are alike, therefore, it suffices to discuss just the case $\ell =1$. The plots corresponding to the sections of the surface $|a_1(\varepsilon,q)|^2$ by families of the planes $|a_1|^2 = const$ and $q=const$ are shown in Figs.~\ref{fig:F1},~\ref{fig:F2}. It is seen from the figures that the closer to the singular point, the steeper and narrower the profile $|a_\ell(\varepsilon,q)|^2$. Eventually, at the singular point itself, it collapses into a straight vertical line connecting the points $|a_\ell|^2=0$ and $|a_\ell|^2=1$.


Any trajectory approaching the singular point in the plane $(\varepsilon,q)$ is a 2D projection of a 3D line, lying on the surface $|a_\ell(\varepsilon,q)|^2$. All these lines end up in the corresponding point of the vertical line with a given single value of $|a_\ell|^2$. However, since the line is \emph{vertical}, all these points are projected onto one and the same singular point in the plane $|a_1(\varepsilon,q)|^2$.

Thus, \emph{the limit depends on the trajectory along which one approaches the singular point.} We keep it in mind and proceed with the analysis.
Let us select any resonant trajectory, take on it a point \mbox{$\varepsilon = \varepsilon_\ell$}, \mbox{$q=q_\ell$}, and consider small departures from this point: \mbox{$\varepsilon = \varepsilon_\ell + \delta\varepsilon$} and \mbox{$q=q_\ell+\delta q$}. Then, expanding the right-hand-sides of Eqs.~\eqref{eq:F_ell_q}, \eqref{eq:G_ell_q} in small $\delta\varepsilon$, $\delta q$, keeping just the leading terms, and bearing in mind that $\varepsilon_\ell \approx -1$ so that in the final expressions it may be replaced by $-1$, we obtain:

At $\ell=1$
\begin{equation}\label{eq:a10}
  a_1 \cong \frac{\pi  q_1^2}{\pi  q_1^2-2 i \left(\delta \varepsilon -2 [q_1 \log q_1]\delta q \right)}.
\end{equation}

At $\ell >1$
\begin{equation}\label{eq:a_ell0}
\!\!\!\!a_\ell \cong \frac{\pi  \ell  \left(\ell ^2-1\right) q_\ell^{2 \ell }}{\pi  \ell  \left(\ell ^2-1\right) q_\ell^{2 \ell }-i 2^{2 \ell -1} (\ell !)^2 [\delta \varepsilon  \left(\ell ^2-1\right)+ 2 q_\ell\delta q]}.
\end{equation}

Now, recall that the limit, we are interested in, depends on the trajectory. Thus, among the diversity of possible trajectories, we must select the correct ones. Based on the physical grounds of the problem, the only correct trajectory for each $\ell$ is the one with a fix $R$ and varying $\omega$, corresponding to the irradiation of a given cylinder by a pulse of a given spectral structure. Moving along this trajectory, we can calculate the integral (over the entire spectrum of the source) partial cross section $\sigma_{\rm sca}^{(\ell)\;{\rm int}}$; then, summarizing them, to find the total cross section $\sigma_{\rm sca}^{\rm tot}(R)$; and, finally, consider the limit $R \rightarrow 0$.

To this end, first, we have to change the independent variables in Eqs.~\eqref{eq:a10}--\eqref{eq:a_ell0} according to the relations \mbox{$\delta q \cong (\partial q/\partial \omega)\delta\omega \equiv (R/c)\delta\omega$} and \mbox{$\delta\varepsilon \cong (\partial\varepsilon/\partial \omega)_{\omega_0}\delta\omega$}, where $\varepsilon(\omega_0)=-1$. Then, Eqs.~\eqref{eq:a10}--\eqref{eq:a_ell0} give rise to the usual Lorentzian profile for $|a_\ell|^2$:
\begin{equation}\label{eq:Lorentzian}
  |a_\ell|^2 \cong \frac{(\Gamma_\ell/2)^2}{(\delta\omega)^2+(\Gamma_\ell/2)^2},
\end{equation}
where
\begin{equation}\label{eq:Gamma_ell}
  \Gamma_\ell \cong \frac{\pi}{(\ell-1)!\ell!(\partial\varepsilon/\partial\omega)_{\omega_0}} \left(\frac{R\omega_0}{2c}\right)^{\!\!2\ell},
\end{equation}
(it is regarded that $0!=1$).

The rest is trivial. According to Eq.~\eqref{eq:Gamma_ell} all $\Gamma_\ell \rightarrow 0$ at $R \rightarrow 0$. On the other hand, $2|a_\ell|^2/(\pi\Gamma_\ell)$, where $|a_\ell|^2$ is given by Eq.~\eqref{eq:Lorentzian}, converges to $\delta$-function at $\Gamma_\ell \rightarrow 0$. Thus, no matter, how narrow the source line is, at small enough $R$ it is much broader than all resonant lines. In this case with respect to the excitation of the resonant modes, the incident radiation should be regarded as incoherent. Then, to get the integral intensity of the radiation scattered by a given multiple, we have to integrate the partial scattering intensities over the spectrum. Accordingly, the integral partial scattering cross section is
\begin{eqnarray}
  \!\!\!\!\! \sigma_{\rm sca}^{(\ell)\;{\rm int}}  & \equiv & \frac{1}{I}\int \sigma_{\rm sca}^{(\ell)} S_\omega d\omega \equiv \int \frac{2\pi^2\Gamma_\ell c}{\omega I}S_\omega\frac{2|a_\ell|^2}{(\pi\Gamma_\ell)} d\omega\nonumber \\
   & \cong &\int \frac{2\pi^2c\Gamma_\ell}{\omega I}S_\omega\delta(\omega-\omega_\ell)d\omega  \cong \frac{2\pi^2 c S_{\omega_0}\Gamma_\ell}{\omega_0 I}.\label{eq:sigma_ell_int}
\end{eqnarray}
Here $S_{\omega}$ and $I\equiv \int S_{\omega}d\omega$ are the spectral and total energy flux densities, respectively.

Next,
\begin{align}
 & \sigma_{\rm sca}^{\rm tot}(R) = \sigma_0\sum_{\ell=1}^{\infty}\frac{1}{(\ell-1)!\ell!} \left(\frac{R\omega_0}{2c}\right)^{\!\!2\ell}\label{eq:sigma_tot}\\
 & < \sigma_0 \sum_{\ell=1}^{\infty}\frac{1}{\ell!} \left(\frac{R\omega_0}{2c}\right)^{\!\!2\ell}=\sigma_0\left(\exp\left[{\left(\frac{R\omega_0}{2c}\right)^2}\right]-1\right),\nonumber
\end{align}
where $\sigma_0 \equiv {4\pi^3 cS_{\omega_0}}/\big[\omega_0 I (\partial\varepsilon/\partial\omega)_{\omega_0}\big]$. Thus, \mbox{$\sigma_{\rm sca}^{\rm tot}(R) \rightarrow 0$} at \mbox{$R \rightarrow 0$,} in agreement with the common sense.


\emph{Conclusions.} We have shown that the paradox of the divergence of the resonant scattering cross section of a cylinder with $\varepsilon \rightarrow -1$ and $R \rightarrow 0$ is resolved if one takes into account that at this point the partial scattering cross sections with $\ell \neq 1$ have singularities. In this case to get the correct result, one, first, has to extend the problem formulation to the scattering of nonmonochromatic light with a finite linewidth; to calculate the total cross section for a small but finite radius of the cylinder and only after that to send the radius to zero. The application of this approach results in the physically reasonable vanishing cross section at $R \rightarrow 0$.

These results show that a subwavelength cylinder with $\varepsilon \approx -1$ hardly can be employed as a superscatterer. However, simultaneously, they create opportunities to observe new interesting effects, coming into been at small but finite $R$. They are associated with different positions of the centers of the resonant lines of various multipoles with respect to the center of the line of the source and different mutual scales of the corresponding linewidths. In this case, an increase in $R$ may push a resonant line outside the source line, which may give rise to a sharp drop in the scattering. This phenomenon should affect both the far field zone (unusual, nonmonotonic dependence of the cross section on $R$) and in the near field zone (sharp changes in the topological structure of the field coursed by small variations in $R$). Despite, these effects lie beyond the scope of the present paper, we believe the present contribution will inspire their study in the very near future.

\begin{acknowledgments}
The authors are grateful to A.E. Miroshnichenko for the discussion of this work and valuable comments. M.I.T. acknowledges the financial support of the Russian Foundation for Basic Research (Grant No. 17-02-00401)
for the problem formulation and analytical study, and the MEPhI Academic Excellence Project (agreement with the Ministry of Education and Science of the Russian Federation of August 27, 2013; project no. 02.a03.21.0005) for the computer symbolic calculations.
\end{acknowledgments}

\bibliography{Cylinder} 

\begin{thebibliography}{11}%
\makeatletter
\providecommand \@ifxundefined [1]{%
 \@ifx{#1\undefined}
}%
\providecommand \@ifnum [1]{%
 \ifnum #1\expandafter \@firstoftwo
 \else \expandafter \@secondoftwo
 \fi
}%
\providecommand \@ifx [1]{%
 \ifx #1\expandafter \@firstoftwo
 \else \expandafter \@secondoftwo
 \fi
}%
\providecommand \natexlab [1]{#1}%
\providecommand \enquote  [1]{``#1''}%
\providecommand \bibnamefont  [1]{#1}%
\providecommand \bibfnamefont [1]{#1}%
\providecommand \citenamefont [1]{#1}%
\providecommand \href@noop [0]{\@secondoftwo}%
\providecommand \href [0]{\begingroup \@sanitize@url \@href}%
\providecommand \@href[1]{\@@startlink{#1}\@@href}%
\providecommand \@@href[1]{\endgroup#1\@@endlink}%
\providecommand \@sanitize@url [0]{\catcode `\\12\catcode `\$12\catcode
  `\&12\catcode `\#12\catcode `\^12\catcode `\_12\catcode `\%12\relax}%
\providecommand \@@startlink[1]{}%
\providecommand \@@endlink[0]{}%
\providecommand \url  [0]{\begingroup\@sanitize@url \@url }%
\providecommand \@url [1]{\endgroup\@href {#1}{\urlprefix }}%
\providecommand \urlprefix  [0]{URL }%
\providecommand \Eprint [0]{\href }%
\providecommand \doibase [0]{http://dx.doi.org/}%
\providecommand \selectlanguage [0]{\@gobble}%
\providecommand \bibinfo  [0]{\@secondoftwo}%
\providecommand \bibfield  [0]{\@secondoftwo}%
\providecommand \translation [1]{[#1]}%
\providecommand \BibitemOpen [0]{}%
\providecommand \bibitemStop [0]{}%
\providecommand \bibitemNoStop [0]{.\EOS\space}%
\providecommand \EOS [0]{\spacefactor3000\relax}%
\providecommand \BibitemShut  [1]{\csname bibitem#1\endcsname}%
\let\auto@bib@innerbib\@empty
\bibitem [{\citenamefont {Grigorenko}\ \emph {et~al.}(2005)\citenamefont
  {Grigorenko}, \citenamefont {Geim}, \citenamefont {Gleeson}, \citenamefont
  {Zhang}, \citenamefont {Firsov}, \citenamefont {Khrushchev},\ and\
  \citenamefont {Petrovic}}]{Grigorenko2005}%
  \BibitemOpen
  \bibfield  {author} {\bibinfo {author} {\bibfnamefont {A.~N.}\ \bibnamefont
  {Grigorenko}}, \bibinfo {author} {\bibfnamefont {A.~K.}\ \bibnamefont
  {Geim}}, \bibinfo {author} {\bibfnamefont {H.~F.}\ \bibnamefont {Gleeson}},
  \bibinfo {author} {\bibfnamefont {Y.}~\bibnamefont {Zhang}}, \bibinfo
  {author} {\bibfnamefont {A.~A.}\ \bibnamefont {Firsov}}, \bibinfo {author}
  {\bibfnamefont {I.~Y.}\ \bibnamefont {Khrushchev}}, \ and\ \bibinfo {author}
  {\bibfnamefont {J.}~\bibnamefont {Petrovic}},\ }\href
  {https://doi.org/10.1038/nature04242} {\bibfield  {journal} {\bibinfo
  {journal} {Nature}\ }\textbf {\bibinfo {volume} {438}},\ \bibinfo {pages}
  {335} (\bibinfo {year} {2005})}\BibitemShut {NoStop}%
\bibitem [{\citenamefont {Miroshnichenko}\ \emph {et~al.}(2010)\citenamefont
  {Miroshnichenko}, \citenamefont {Flach},\ and\ \citenamefont
  {Kivshar}}]{Miroshnichenko:RMP:2010}%
  \BibitemOpen
  \bibfield  {author} {\bibinfo {author} {\bibfnamefont {A.}~\bibnamefont
  {Miroshnichenko}}, \bibinfo {author} {\bibfnamefont {S.}~\bibnamefont
  {Flach}}, \ and\ \bibinfo {author} {\bibfnamefont {Y.}~\bibnamefont
  {Kivshar}},\ }\href {\doibase 10.1103/RevModPhys.82.2257} {\bibfield
  {journal} {\bibinfo  {journal} {Rev. Mod. Phys.}\ }\textbf {\bibinfo {volume}
  {82}},\ \bibinfo {pages} {2257} (\bibinfo {year} {2010})}\BibitemShut
  {NoStop}%
\bibitem [{\citenamefont {Schwartz}\ \emph {et~al.}(2009)\citenamefont
  {Schwartz}, \citenamefont {Shetty}, \citenamefont {Price}, \citenamefont
  {Stafford}, \citenamefont {Wang}, \citenamefont {Uthamanthil}, \citenamefont
  {Pham}, \citenamefont {McNichols}, \citenamefont {Coleman},\ and\
  \citenamefont {Payne}}]{schwartz2009feasibility}%
  \BibitemOpen
  \bibfield  {author} {\bibinfo {author} {\bibfnamefont {J.~A.}\ \bibnamefont
  {Schwartz}}, \bibinfo {author} {\bibfnamefont {A.~M.}\ \bibnamefont
  {Shetty}}, \bibinfo {author} {\bibfnamefont {R.~E.}\ \bibnamefont {Price}},
  \bibinfo {author} {\bibfnamefont {R.~J.}\ \bibnamefont {Stafford}}, \bibinfo
  {author} {\bibfnamefont {J.~C.}\ \bibnamefont {Wang}}, \bibinfo {author}
  {\bibfnamefont {R.~K.}\ \bibnamefont {Uthamanthil}}, \bibinfo {author}
  {\bibfnamefont {K.}~\bibnamefont {Pham}}, \bibinfo {author} {\bibfnamefont
  {R.~J.}\ \bibnamefont {McNichols}}, \bibinfo {author} {\bibfnamefont {C.~L.}\
  \bibnamefont {Coleman}}, \ and\ \bibinfo {author} {\bibfnamefont {J.~D.}\
  \bibnamefont {Payne}},\ }\href@noop {} {\bibfield  {journal} {\bibinfo
  {journal} {Cancer research}\ }\textbf {\bibinfo {volume} {69}},\ \bibinfo
  {pages} {1659} (\bibinfo {year} {2009})}\BibitemShut {NoStop}%
\bibitem [{\citenamefont {Tribelsky}\ and\ \citenamefont
  {Fukumoto}(2016)}]{Tribelsky2016}%
  \BibitemOpen
  \bibfield  {author} {\bibinfo {author} {\bibfnamefont {M.}~\bibnamefont
  {Tribelsky}}\ and\ \bibinfo {author} {\bibfnamefont {Y.}~\bibnamefont
  {Fukumoto}},\ }\href {\doibase 10.1364/BOE.7.002781} {\bibfield  {journal}
  {\bibinfo  {journal} {Biomedical Optics Express}\ }\textbf {\bibinfo {volume}
  {7}} (\bibinfo {year} {2016}),\ 10.1364/BOE.7.002781}\BibitemShut {NoStop}%
\bibitem [{\citenamefont {Tribelsky}\ \emph {et~al.}(2015)\citenamefont
  {Tribelsky}, \citenamefont {Geffrin}, \citenamefont {Litman}, \citenamefont
  {Eyraud},\ and\ \citenamefont {Moreno}}]{Tribelsky2015}%
  \BibitemOpen
  \bibfield  {author} {\bibinfo {author} {\bibfnamefont {M.}~\bibnamefont
  {Tribelsky}}, \bibinfo {author} {\bibfnamefont {J.-M.}\ \bibnamefont
  {Geffrin}}, \bibinfo {author} {\bibfnamefont {A.}~\bibnamefont {Litman}},
  \bibinfo {author} {\bibfnamefont {C.}~\bibnamefont {Eyraud}}, \ and\ \bibinfo
  {author} {\bibfnamefont {F.}~\bibnamefont {Moreno}},\ }\href {\doibase
  10.1038/srep12288} {\bibfield  {journal} {\bibinfo  {journal} {Scientific
  Reports}\ }\textbf {\bibinfo {volume} {5}} (\bibinfo {year} {2015}),\
  10.1038/srep12288}\BibitemShut {NoStop}%
\bibitem [{\citenamefont {Klimov}(2014)}]{Klimov:Book:2014}%
  \BibitemOpen
  \bibfield  {author} {\bibinfo {author} {\bibfnamefont {V.}~\bibnamefont
  {Klimov}},\ }\href@noop {} {\emph {\bibinfo {title} {Nanoplasmonics}}}\
  (\bibinfo  {publisher} {Pan Stanford},\ \bibinfo {year} {2014})\BibitemShut
  {NoStop}%
\bibitem [{\citenamefont {Ruan}\ and\ \citenamefont {Fan}(2010)}]{Ruan2010}%
  \BibitemOpen
  \bibfield  {author} {\bibinfo {author} {\bibfnamefont {Z.}~\bibnamefont
  {Ruan}}\ and\ \bibinfo {author} {\bibfnamefont {S.}~\bibnamefont {Fan}},\
  }\href {\doibase 10.1103/PhysRevLett.105.013901} {\bibfield  {journal}
  {\bibinfo  {journal} {Physical Review Letters}\ }\textbf {\bibinfo {volume}
  {105}},\ \bibinfo {pages} {1} (\bibinfo {year} {2010})}\BibitemShut {NoStop}%
\bibitem [{\citenamefont {Tribelsky}\ and\ \citenamefont
  {Luk'yanchuk}(2006)}]{Tribelsky2006}%
  \BibitemOpen
  \bibfield  {author} {\bibinfo {author} {\bibfnamefont {M.}~\bibnamefont
  {Tribelsky}}\ and\ \bibinfo {author} {\bibfnamefont {B.}~\bibnamefont
  {Luk'yanchuk}},\ }\href {\doibase 10.1103/PhysRevLett.97.263902} {\bibfield
  {journal} {\bibinfo  {journal} {Physical Review Letters}\ }\textbf {\bibinfo
  {volume} {97}},\ \bibinfo {pages} {263902} (\bibinfo {year}
  {2006})}\BibitemShut {NoStop}%
\bibitem [{\citenamefont {Bohren}\ and\ \citenamefont
  {Huffman}(2008)}]{bohren2008absorption}%
  \BibitemOpen
  \bibfield  {author} {\bibinfo {author} {\bibfnamefont {C.~F.}\ \bibnamefont
  {Bohren}}\ and\ \bibinfo {author} {\bibfnamefont {D.~R.}\ \bibnamefont
  {Huffman}},\ }\href@noop {} {\emph {\bibinfo {title} {Absorption and
  scattering of light by small particles}}}\ (\bibinfo  {publisher} {John Wiley
  \& Sons},\ \bibinfo {year} {2008})\BibitemShut {NoStop}%
\bibitem [{\citenamefont {Luk'yanchuk}\ \emph {et~al.}(2007)\citenamefont
  {Luk'yanchuk}, \citenamefont {Tribelsky}, \citenamefont {Ternovsky},
  \citenamefont {Wang}, \citenamefont {Hong}, \citenamefont {Shi},\ and\
  \citenamefont {Chong}}]{lukyanchuk2007peculiarities931101}%
  \BibitemOpen
  \bibfield  {author} {\bibinfo {author} {\bibfnamefont {B.}~\bibnamefont
  {Luk'yanchuk}}, \bibinfo {author} {\bibfnamefont {M.}~\bibnamefont
  {Tribelsky}}, \bibinfo {author} {\bibfnamefont {V.}~\bibnamefont
  {Ternovsky}}, \bibinfo {author} {\bibfnamefont {Z.}~\bibnamefont {Wang}},
  \bibinfo {author} {\bibfnamefont {M.}~\bibnamefont {Hong}}, \bibinfo {author}
  {\bibfnamefont {L.}~\bibnamefont {Shi}}, \ and\ \bibinfo {author}
  {\bibfnamefont {T.}~\bibnamefont {Chong}},\ }\href {\doibase
  10.1088/1464-4258/9/9/S03} {\bibfield  {journal} {\bibinfo  {journal}
  {Journal of Optics A: Pure and Applied Optics}\ }\textbf {\bibinfo {volume}
  {9}},\ \bibinfo {pages} {S294} (\bibinfo {year} {2007})}\BibitemShut
  {NoStop}%
\bibitem [{Note1()}]{Note1}%
  \BibitemOpen
  \bibinfo {note} {The possibility to transform $\protect \mathfrak {F}_\ell $
  and $\protect \mathfrak {G}_\ell $ in Eq.~\protect \textup {\hbox
  {\mathsurround \z@ \protect \normalfont (\ignorespaces \ref {eq:a_ell}\unskip
  \@@italiccorr )}} into purely real $F_\ell $ and $G_\ell $ at purely real
  $\varepsilon $ follows from Eq.~\protect \textup {\hbox {\mathsurround \z@
  \protect \normalfont (\ignorespaces \ref {eq:F_ell_G_ell}\unskip
  \@@italiccorr )}} and is not related to the smallness of $q$.}\BibitemShut
  {Stop}%
\end{thebibliography}%

\end{document}